\documentclass[preprint,aps,prl]{revtex4-1}

\usepackage{graphicx}
\usepackage{dcolumn}
\usepackage{bm}

\usepackage{epstopdf}
\usepackage{longtable}

\begin{document}
\large
\begin{center}
{\bf Braiding of Abelian and Non-Abelian Anyons\\
 in the Fractional Quantum Hall Effect}
\end{center}

\normalsize

\vspace*{0.1in}

\begin{center}
Sanghun An$^{1}$,  P. Jiang$^{1,2}$, H. Choi$^{1}$, W. Kang$^{1}$, S.H. Simon$^{3}$, L.N. Pfeiffer$^{4}$,  K.W. West$^{4}$, and K.W. Baldwin$^{4}$\\

{\it \small
$^{1}$James Franck Institute and Department of Physics,  University of Chicago, Chicago, Illinois 60637\\
$^{2}$Department of Physics, National Taiwan Normal University, Taipei, Taiwan\\
$^{3}$Rudolf Peierls Centre for Theoretical Physics, 1 Keble Road, Oxford University, OX1 3NP, UK\\
$^{4}$Department of Electrical Engineering, Princeton University, Princeton, NJ 08544\\}

\end{center}

{\bf In two spatial dimensions quantum mechanical particles are not limited to being bosons or fermions as they are in three dimensions, but can be particles known as anyons\cite{Leinaas77,Moore91,Nayak08} which obey braiding statistics ---  meaning that the quantum mechanical state of a system is altered in a particular way when one anyon is moved (braided) around another, independent of the specific path chosen. Such anyons come in two major varieties --- Abelian and non-Abelian --- both of which were long ago predicted\cite{Halperin84,Arovas84,Moore91,Nayak08} to be realized in certain Fractional Quantum Hall (FQH) systems.  Non-Abelian anyons have recently received considerable attention as a promising platform for topological quantum computation\cite{Kitaev03, Freedman03, DasSarma05, Nayak08}.     However, experimental demonstration of anyonic braiding properties has remained elusive and very controversial\cite{Camino05,Willett10}.   In this paper, we report on the study of Abelian and non-Abelian statistics through Fabry-Perot interferometry.  Our detection of phase slips in quantum interference experiments demonstrates a powerful, new way of detecting braiding of anyons.  We confirm the Abelian anyonic braiding statistics in the $\nu = 7/3$ FQH state through detection of the predicted statistical phase angle of $2\pi/3$, consistent with a change of the anyonic particle number by one.   The $\nu = 5/2$ FQH state\cite{Willett87} is theoretically believed to harbor non-Abelian anyons\cite{Moore91,Nayak08} which are Majorana, meaning that each pair of quasiparticles contain a neutral fermion orbital which can be occupied or unoccupied and hence can act as a qubit.  In this case our observed statistical phase slips agree with a theoretical model\cite{RosenowSimon} where the Majoranas are strongly coupled to each other, and strongly coupled to the edge modes of the interferometer. In particular, an observed  phase slip of approximately $\pi$ is interpreted as a sudden flip of a qubit, or entry of a neutral fermion into the interferometer. Our results provide compelling support for the existence of non-Abelian anyons.} 

\pagebreak

The most promising systems for the realization of anyons are the FQH states in two-dimensional electron systems at low temperatures and high magnetic fields.
Each FQH state is labeled by its filling fraction $\nu$,  the dimensionless ratio of electron density to magnetic flux.   The primary FQH states at $\nu  = 1/m$, with $m$ odd, are thought to be an imcompressible quantum fluid.
Theory predicts\cite{Halperin84,Arovas84} that the quasiparticle excitations of these states should be anyons with fractional charge $q = e/m$ and fractional statistical phase  $\theta = \pi/m$ occurring whenever two such anyons are exchanged in a clockwise manner. While there is convincing experimental evidence in support of the fractional charge\cite{dePicciotto97,Saminadayar97,Dolev08,Venkatachalam11}, experiments on the study of fractional statistics have been received with skepticism\cite{Camino05}.

The FQH state at $\nu = 5/2$ is considered to be the most promising system for the demonstration of non-Abelian anyon braiding statistics.   The quasiparticle excitations are expected to be anyons with charge $q = e/4$  and whose braiding statistics have both an Abelian part with $\theta=\pi/8$ and a more complicated non-Abelian part which associates a Majorana with each quasiparticle\cite{Moore91,Nayak08}.

Braiding statistics of anyonic quasiparticles for FQH systems may be detected through quantum interferometry experiments, in which test quasiparticles ``braid" a group of localized quasiparticles at the center of the interferometer\cite{Nayak08,DasSarma05,Camino05,Willett10,Chamon97,Stern06,Bonderson06,Zhang09,Willett09,Ofek10,Choi11}.
In a quantum Hall Fabry-Perot interferometer, interference trajectories are realized through coherent transport along edge channels and tunneling across a pair of split-gated constrictions, which act as beam splitters  (See Fig. 1b).   In the Abelian case, assuming that tunneling at the first and the second constrictions occur with amplitude $t_1$ and $t_2$, the conductance, G, across the interferometer is
\begin{equation}
{\rm G} \propto |t_{1}|^{2} + |t_{2}|^{2} + 2{\rm Re} \{t_{1}^{*}t_{2}e^{i\phi}\}
\end{equation}
The resulting interference phase, $\phi$, is a sum of the Aharonov-Bohm phase due to the magnetic flux enclosed by the edge state trajectory and the statistical phase $2 \theta$ (a loop equals two exchanges, hence $2 \theta$) for each quasiparticle encircled by the edge. The general scheme of this measurement is to observe discrete phase-slip events of $2 \theta$ associated with the sudden entry or exit of a quasiparticle\cite{Grosfeld06,RosenowSimon} from the interferometer.

The Fabry-Perot interferometer was fabricated from a high mobility, symmetrically doped GaAs/AlGaAs quantum well. 
The layout of the interferometer is shown in Fig. 1a.  The diameter of the interferometer is 1.2$ \mu$m, and the width of the two constrictions is $\sim$400 nm. The sample was mounted on a dilution refrigerator capable of reaching below 10 mK.  [Magnetotransport through the interferometer is shown in the supplementary material.]
The diagonal resistance, R$_D$, is measured as the voltage bias on the plunger gate (lower middle gate of Fig. 1a) is swept at a steady rate. The bias changes the encircled area resulting in interference oscillations. Experimental studies of interference in the integer and fractional quantum Hall regimes have been reported earlier\cite{Camino05,Zhang09,Willett09,Willett10,Ofek10,Choi11}.  The effect of Coulomb charging in the interferometer may be either strong or weak, in which case we say the device is in either the Coulomb-dominated or Aharonov-Bohm regime respectively\cite{Rosenow07,Halperin11,Zhang09,Ofek10,Choi11}.  The phases observed (for example the gate period) can be modified by such Coulomb effects if they are present.  However, it has also been discussed\cite{RosenowSimon} that under certain reasonable conditions, the phase slip when a quasiparticle enters the interferometer could be immune to these modifications even in the Coulomb-dominated regime.

Fig.~1 illustrates sweeps of the plunger gate voltage V$_P$ for the $\nu = 2$, $7/3$ and $5/2$ FQH states. In the case of the $\nu = 2$ state (Fig.~1c), R$_D$ oscillates periodically with V$_P$.  For the 7/3 state (Fig. 1d), R$_D$ shows a periodic behavior for V$_P < -27$ mV.  For  $-27$ mV  $< $V$_P < -18$mV, the behavior shows a telegraph noise behavior between two curves (black and blue) that are out of phase by exactly $\Delta \phi=2 \pi/3$.   This noise reflects the phase associated with an anyon randomly entering and exiting the interferometer.   Another $2\pi/3$ phase slip can be identified near $V_P = -15$mV [See Supplementary Material for numerical analysis].   Observation of these phase slips demonstrates the predicted braiding statistics of the $e/3$ anyons for the 7/3 FQH state.

For the 5/2 state, the above equation for the interference may be modified to account for the non-Abelian properties of the quasiparticle excitations\cite{DasSarma05,Nayak08,Fradkin98,Stern06,Bonderson06}.   In this case, if there is an even number of quasiparticles in the interferometer, interference should be observed, but the phase of the interference may be shifted by $\pi$ depending on the parity of the number of neutral fermion orbitals that are occupied inside the interferometer (i.e., the settings of the qubits).   A flip of a qubit can suddenly occur (equivalent to a neutral fermion entering the device) resulting in a $\pi$ phase slip in the interference\cite{RosenowSimon}.

If there are an odd number of $e/4$ quasiparticles inside the interferometer, no interference is predicted so long as the quasiparticles are isolated from the edge\cite{Stern06,Bonderson06}.    However, for a small enough device, quasiparticles will always be strongly coupled to the edge.  We believe that our device is in this limit.  In this case\cite{RosenowSimon,Rosenow09,BisharaNayak}, instead, it is predicted that the non-Abelian part of the quasiparticle should be absorbed into the edge;  interference does not vanish, and one observes only the phase $2 \theta= \pi/4$ from the Abelian part.   Thus the full prediction\cite{RosenowSimon} for this device is that one should observe $\pi$ phase slips associated with flips of the qubit, and $\pi/4$ slips associated with quasiparticle addition (both can occur together resulting in a slip of $5 \pi/4$).

Fig. 1e illustrates a sweep of the plunger gate voltage and the analysis of the phase slips for the 5/2 FQH state. Somewhat similar to the 7/3 data,  R$_D$ shows an approximate periodic behavior with a clear phase slip occurring around -35 mV.    The measured phase slip angle is very close to  $\Delta\phi = 1.25\pi$ in agreement with theoretical prediction.

Fig. 2a shows the detailed interference profile of the $\nu = 5/2$ FQHE state for many plunger gate sweeps at 21 successive values of magnetic fields.    (The general structure of the plot indicates that this device is Coulomb-dominated).    A striking feature is that R$_D$ does not change monotonically with gate voltage and magnetic field. There are instances where a minimum becomes a maximum and vice versa (i.e., phase slips of approximately $\pi$).

In Figs. 2b-2f show representative plunger gate voltage sweeps from Fig. 2a.
Fit of the R$_D$ sweeps show that an overwhelming majority of the sweeps exhibit clear phase slips of $\Delta\phi \approx \pi $ to $5\pi /4$ in the range of gate voltages between -32 and -37 mV.

In order to make a unbiased analysis of these phase slip angles, we have developed a computerized algorithm to best determine the phase slips present in Fig. 2a.  (See Supplementary Material.)   At each point in the $(V_P, B)$ plane, the gate sweep is fitted with two sine wave segments, one for $V > V_P$ and one for $V < V_P$, and the relative phase slip between these two curves is plotted in Fig 3a.  The red background represents the phase slips near 0 and $2\pi$ (meaning no phase slip).     The nontrivial regions of this plot are then histogrammed in Fig 3b.  Two prominent persistent features are observed: most clearly, strong phase slips ranging from $\pi$ to $5\pi/4$ are observed in the range of gate voltages from -42 to -31 mV (blue to light blue).  Phase slips near $.35\pi$ (yellow-green) are observed between -27 to -17 mV.  Much more weakly, both the yellow-green and blue regions show slips near $0.7\pi$.

Given that our system is Coulomb-dominated, the theoretically expected phase slip of $\pi/4$ does not have to hold precisely\cite{RosenowSimon}, and it is not unreasonable to conclude that the measured $0.35 \pi$ is a reflection of a $\pi/4$ slip with some modification. The weak $0.7 \pi$ peak could be a pair of $0.35 \pi$ slips in a row.   It is also possible that the slips between $\pi$ and $5 \pi/4$ are one or the other value and these are either modified slightly by Coulomb effects, or are simply measured imperfectly due to other noise in our data.   Whether or not we can pin down the expected $\pi/4$ and $5\pi/4$, it is quite clear that slips very near $\pi$ are extremely dominant.  While it has been pointed out\cite{Ilan} that interferometry experiments may not uniquely identify a particular quantum Hall state, such slips are clear evidence of neutral fermions, and are a crucial step towards demonstrating the  reality of non-Abelian anyons.

In summary, we have detected the braiding statistics of anyons.   For the $\nu=7/3$ case our data clearly demonstrates the predicted braiding statistics.  For $\nu=5/2$ our data is in agreement with current theory, clearly observing a neutral fermion.
Finally, we note that it is quite difficult to obtain data of this type since too much telegraph noise can easily make interpretation impossible, and too little removes the desired slips.
Nonetheless, with persistence this is a powerful technique for probing the anyonic properties of FQH states.
We anticipate that the fidelity of anyonic detection will improve with time, possibly leading to the realization of a topological quantum information processor.

\vspace*{0.05in}
We thank B. Rosenow, I. Gruzberg, and P. Wiegmann for discussions. S.A., P.J., H.C., and W.K. acknowledge support by Microsoft Project Q and NSF MRSEC Program through the University of Chicago Materials Center (DMR-0820054). S.H.S. was supported by EPSRC grants EP/I032487/1 and EP/I031014/1.  L.N.P, K.W.W., and K.W.B acknowledge partial support from the Gordon and Betty Moore Foundation and the NSF MRSEC Program through the Princeton Center for Complex Materials (DMR-0819860).\

\pagebreak

\includegraphics[width=5.5in]{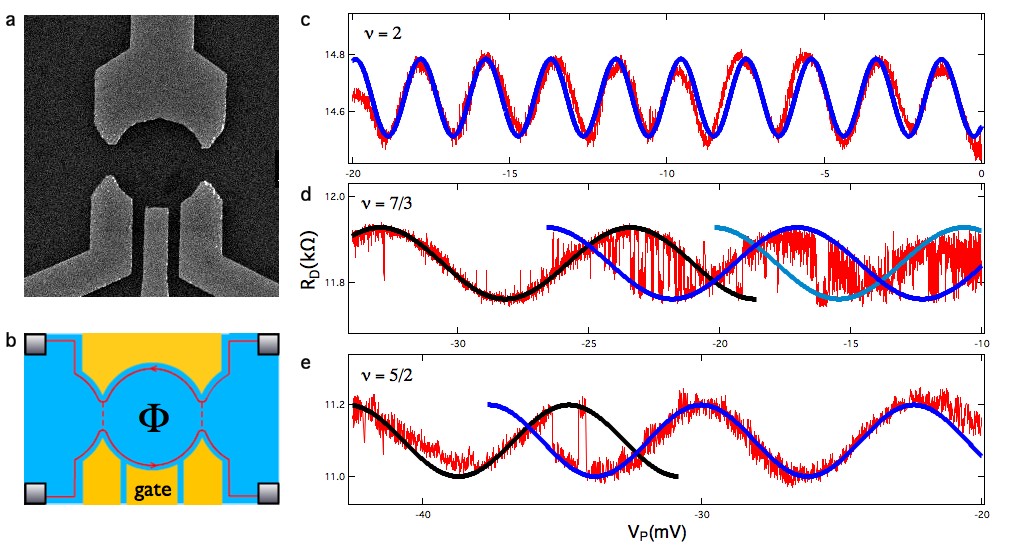}
\noindent
Figure 1: Fabry-Perot interferometer and interference measurements.\\
{\bf a}, A SEM view of the Fabry-Perot interferometer (FPI). The low temperature mobility of the unprocessed sample was $2.83\times 10^{7}$cm$^{2}$/Vs with an electron density of $3.3 \times 10^{11}$ cm$^{-2}$. The interferometer  was initially defined using e-beam lithography, dry etched to the depth of the two-dimensional electron system, and  metallized with TiAu prior to lift-off. The diameter of the interferometer is 1.2$ \mu$m, and the width of the two constrictions is $\sim$400 nm.  {\bf b}, Expected edge state trajectory (red) in the quantum Hall regime for the FPI. Two narrow constrictions serve as beam splitters and interfere particles tunneling at the two constrictions. The observed interference phase is a sum of the Aharonov-Bohm phase due to the enclosed magnetic flux $\Phi$ and the statistical phase due to the exchange. {\bf c}, Diagonal resistance, R$_D$, as a function of the plunger gate voltage at the $\nu = 2$ quantum Hall state under B = 7.5614 tesla. Bias of the side gate from {\bf a} changes the area enclosed by interfering particles and leads to a periodic oscillation. 
{\bf d}, Measurement  of R$_D$ at the $\nu = 7/3$ fractional quantum Hall state under B = 5.7139 tesla. In conjunction with phase slips of $\Delta\phi = 2\pi/3$, telegraph noise is observed in some regions of gate voltage. Telegraph noise is most pronounced between -20 and 25mV, but telegraph noise behavior is seen above -20mV as well, presumably from discrete switching of the number of quasiparticles in the interferometer.   The black curve is the best fit of the large bias portion of the data below -20mV. The blue curve is the same curve as the black curve but phase-shifted by $2\pi/3$. The light blue is further phase shifted by $4\pi/3$ relative to the black curve. With the phase slips occurring, the data can be parameterized from the black to blue to the light blue curves with the successive phase slips that are very close to $2\pi/3$. {\bf e}, Measurement  of R$_D$ at the $\nu = 5/2$ fractional quantum Hall state under B = 5.5275 tesla.  The blue curve is the same curve as the black curve but phase-shifted by $5\pi/4$. Phase slip events occur between -35.5 and -34 mV.

\vspace{0.1in}

\includegraphics[width=5.5in]{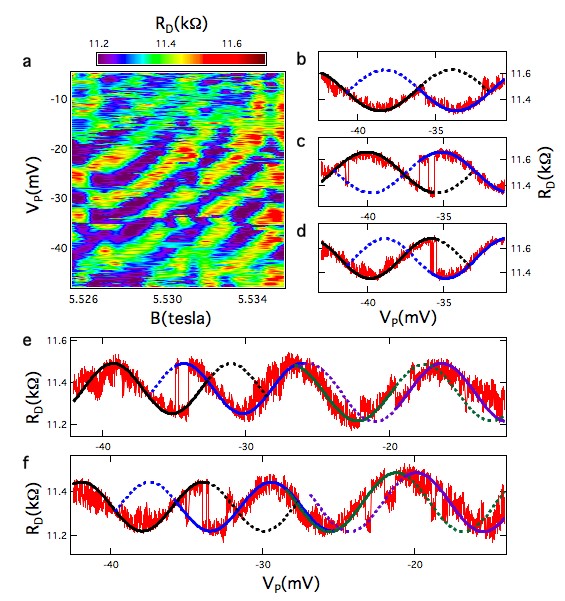}

\noindent
Figure 2: Diagonal resistance and phase slips at the $\nu = 5/2$ fractional quantum Hall state.\\
{\bf a}, The diagonal resistance, R$_D$, with a smooth background subtracted as a function of magnetic field, B, and plunger gate voltage, V$_{P}$. R$_D$ exhibits a series of oscillations whose minima and maxima tend to evolve with a positive $d$V$_{P}/d$B curvature over the measured ranges of gate voltage and magnetic field. The $d$V$_{P}/d$B curvature places the interference in the Coulomb dominated regime of interferometry where Coulomb interaction plays an important role.
{\bf b-d} Representative phase slips of $\pi$ and $5\pi/4$ between -37 and -30mV. The phase slip of $\sim \pi$ leads to a reversal of the parity of the R$_D$. {\bf b}: 5.5330 telsa, {\bf c}: 5.5340 tesla, {\bf d}: 5.5355 tesla. {\bf e}, An example of sweep under B = 5.5315 tesla where there is a multiple phase slip events with $\Delta\phi \approx \pm5\pi/4$ at -34mV and $\sim \pm\pi/4$ phase slips between -19 and -17mV.  {\bf f}, Plunger voltage sweep under B = 5.5305 tesla with multiple phase slip events with $\Delta\phi \approx  \pm\pi$  between -35 and -33.5mV and $\sim \pm\pi/4$ phase slips between -23 and -18 mV.

\vspace{0.1in}

\includegraphics[width=6.2in]{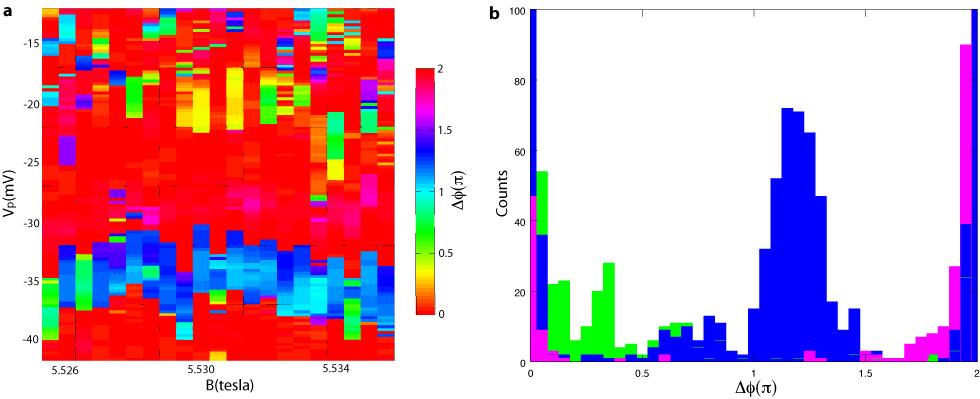}
Figure 3: Phase slips for the $\nu = 5/2$ fractional quantum Hall state.\\
{\bf a}, Plot of phase slip angles as a function of plunger gate bias voltage and  magnetic field from the diagonal resistance data from Fig. 2a. A computerized algorithm was used to determine the best phase slip in an unbiased way.
{\bf b}, Histogram of non-trivial regions of the phase slip angles from Fig. 3a.

\newpage

\begin{center}
 {\bf  Supplementary Material}
\end{center}

\section{Magnetotransport}

Supplementary Fig. 1 compares the Hall resistance, R$_{xy}$, of the bulk, unprocessed part of the sample and the diagonal resistance, R$_D$, through the Fabry-Perot interferometer (FPI) in the second Landau level. These data were taken after an illumination by a red LED. R$_{xy}$ in the bulk part of the sample shows fractional quantum Hall (FQH) states at filling factors $\nu = 5/2, 7/3, 8/3$, and 14/5. In addition, a pair of reenetrant insulating states to $\nu = 2$ and 3 are respectively observed on high and low sides of the $\nu = 5/2$ plateau.  The FQH states within the FPI is noticeably weaker compared to that found in the bulk. The R$_D$ through the interferometer exhibits a noticeable decrease in the electron density and degradation of various quantum Hall states. Only $\nu = 5/2$ and 7/3 states are visible and reentrant insulating states to $\nu = 3$ are suppressed. Such changes may be expected due to the small size of the interferometer and the side-depletion coming from the confinement.

\vspace*{0.25in}

\includegraphics[width=5in]{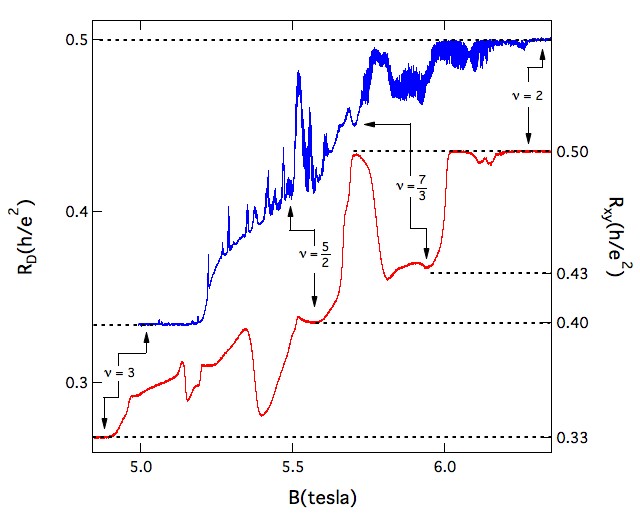}\\
\noindent
{\small Supplementary Figure 1: Hall resistance, R$_{xy}$, of the bulk, two-dimensional electrons and the diagonal resistance, R$_D$, through the Fabry-Perot interferometer}

\section{Reproducibility of Anyonic Phase Slips}

An important question regarding our results of anyonic braiding and phase slips is how reproducible the data is. It turns out that observation of phase slip for both $\nu = 7/3$ and $5/2$ fractional quantum Hall states is rather difficult because it is not easy to realize the condition within the FPI under which phase slips can be detected via DC transport measurements. In order to realize the $\nu = 5/2$ FQH state inside the FPI, we must illuminate the FPI using a LED during the cooldown. Too much or too little illumination can make it hard to detect any interference. Too much light activates low-frequency charge noise, which can overwhelm the interference signals. Too little light leads to low mobility device with no 5/2 state being realized and even no discernible interference signature. 

In our experience in working with many interferometers, whether or not a clear quantum interference can be detected depends critically on the electronic state that is realized within the two narrow constrictions that define the openings of the FPI. It is not always predictable whether or not the constrictions will behave like good beam-splitters. A large density difference between two constrictions seems to be universally bad in suppressing the interference signal. With a reliably good interferometer - where good is defined as being able to detect interference signal and the FQH states in the second Landau level - we aim to optimize the illumination and cooldown procedure to realize uniformity between two constrictions and a low disorder FPI that can sustain robust FQH states. 

Another variability in the experiment is the timescale of the phase slip events, which is expected to be determined by the local potential. 
If the phase slips and the corresponding telegraph noise occurs at too fast of a time scale, then it becomes difficult to detect them via DC transport. If the time scale is too long, then it becomes difficult to collect sufficient number of events within a practical time scale to be feasible. 

With all these factors in play, it is not possible to detect clear phase slip events in interference measurements for every cooldown. However, we have succeeded in detecting phase slips in the $\nu = 5/2$ state on at least three separate cooldowns. In the two out of three cooldowns, we also detected phase slips in the $\nu = 7/3$ state. (In the third cooldown, an equipment failure prevented the search for phase slip events for the $\nu = 7/3$ state.) In Supplementary Figs. 2 and 3 we provide data of phase slips for the $\nu = 7/3$ and $\nu = 5/2$ respectively from the cooldowns \#1 and \#2. The $\nu = 7/3$ phase slip data shown in Fig. 1 of the main paper was taken during the cooldown \#1. The $\nu = 5/2$ phase slip data in the main paper came from the third cooldown. The phase slip angles and the best fit to the data were verified by the computerized algorithm described in Section III.

\vspace*{0.25in}

\includegraphics[width=5in]{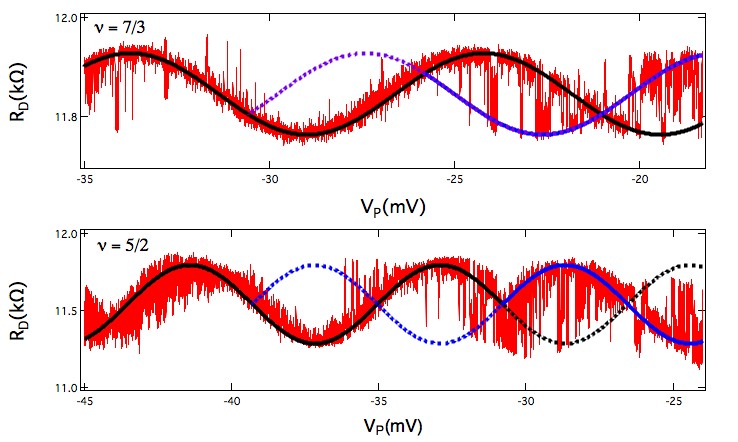}\\
\noindent
{\small Supplementary Figure 2: Phase slips for the $\nu = 7/3$ (upper) and $\nu = 5/2$ (lower) fractional quantum Hall states from cooldown \#1. For the $\nu = 7/3$ state, $\Delta\phi = 2\pi/3$ was observed between -26 and -22mV. For the $\nu = 5/2$ state, $\Delta\phi \approx \pi$ was observed at -31mV. The $\nu = 7/3$ data in the Figure 1 of the paper came from this cooldown.}\\

\vspace*{0.15in}

\includegraphics[width=5in]{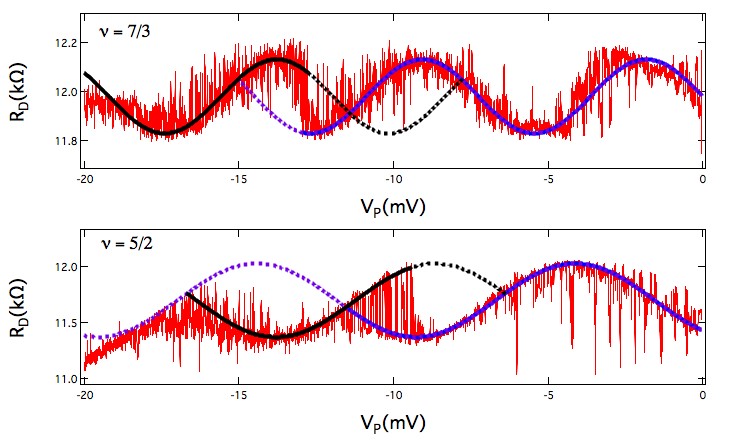}\\
\noindent
{\small Supplementary Figure 3: Phase slips for the $\nu = 7/3$ (upper) and $\nu = 5/2$ (lower) fractional quantum Hall states from cooldown \#2. For the $\nu = 7/3$ state, $\Delta\phi = 2\pi/3$ occurs at -12.5mV. For the $\nu = 5/2$ state, $\Delta\phi \approx \pi$ phase slip occurs multiple times between -12.5 and -9mV.}

\section{Phase slip fitting algorithms}

Slips are identified separately for each sweep of the gate voltage $V_P$.  For each point $V_P^0$ the program maximizes a fit quality of the measured resistance $R_D$ to two model sine-waves.   We propose sine wave ${\cal R}_D^-$ for $V_P < V_P^0$ and sine wave ${\cal R}_D^+$ for $V_P > V_P^0$.  The form of these proposed model functions are 
\begin{eqnarray*}
  {\cal R}_{D}^+(V_P)  &=&   a + b V_P  + c \cos(f V_P - \phi_+) \\
  {\cal R}_{D}^-(V_P)  &=&   a + b V_P  + c \cos(f V_P - \phi_-)
\end{eqnarray*}
where  $a,b,c,f,\phi_+,\phi_-$ are all free parameters (note that a linear drift term $b$ is also included). 

The fit is performed over the range $V_P^0 - \delta V < V_P < V_P^0 + \delta V$ for some fixed range $\delta V$.   The fit quality that is maximized is the following function
$$
Q =
 \int_{V_P^0 - \delta V}^{V_P^0+ \delta V} dV_P  \,\,  F\left[ R_{D}(V_P) \right]
$$
where
$$
F[R_D(V_P)] = \Theta(V_P -V_P^0) G[R_D(V_P) - {\cal R}_D^{+}(V_P) ]  + \Theta(V_P^0 - V_P) G[R_D -(V_P) {\cal R}_D^{-}(V_P) ]
$$
Here the $G$ is a nearness function for the fit
$$
 G[x] = \frac{1}{1 + (x/\delta R)^2}
$$
with $\delta R$ a fixed parameter.    The nearness function is maximized if the model function is exactly equal to the measured value (i.e., for $x=0$).   Essentially this is least squares fit, but the penalty is softened at larger deviation from the model $(|x| \gtrsim \delta R)$ so as not to heavily penalize outlier points which are very far from the model.   Above $\Theta$ plays the role of the Heaviside step function, so that one tries to fit the measurement to the model function ${\cal R}_D^+$ for $V_P > V_P^0$ and to ${\cal R}_D^-$ for $V_P < V_P^0$. For better numerical performance of our algorithm and smoother results $\Theta$ is smeared somewhat and instead takes the form
$$
 \Theta(x) = \left\{  \begin{array}{lcl}  1  & ~~~~& x \geq 0 \\
                                    e^{x/w}  &     &  x < 0  \end{array} \right.
$$
with some fixed smearing distance $w$.   Fixing the three $w$,  $\delta V$, and $\delta R$, the  remaining parameters $a,b,c,f,\phi_+,\phi_-$ are optimized numerically to maximize the quality of the fit $Q$.  Since this is a nonlinear maximization, occasionally the algorithm (Nelder-Mead) will numerically converge to a local rather than global maximum.   To combat this problem, the  algorithm is run ten times with different seed values and the best result is chosen (typically only a few runs will not converge to almost the same result).  To remain unbiased, the phases $\phi_+$ and $\phi_-$ for the seed values are always chosen randomly.

Once the global maximum quality fit is found, the measured phase slip at the point $V_P^0$ is then given by $\phi_+ - \phi_-$.    Results do not differ much for a wide range of values of the three chosen parameters $w, \delta V$ and $\delta R$.

\vspace*{0.25in}

\includegraphics[width=6.15in]{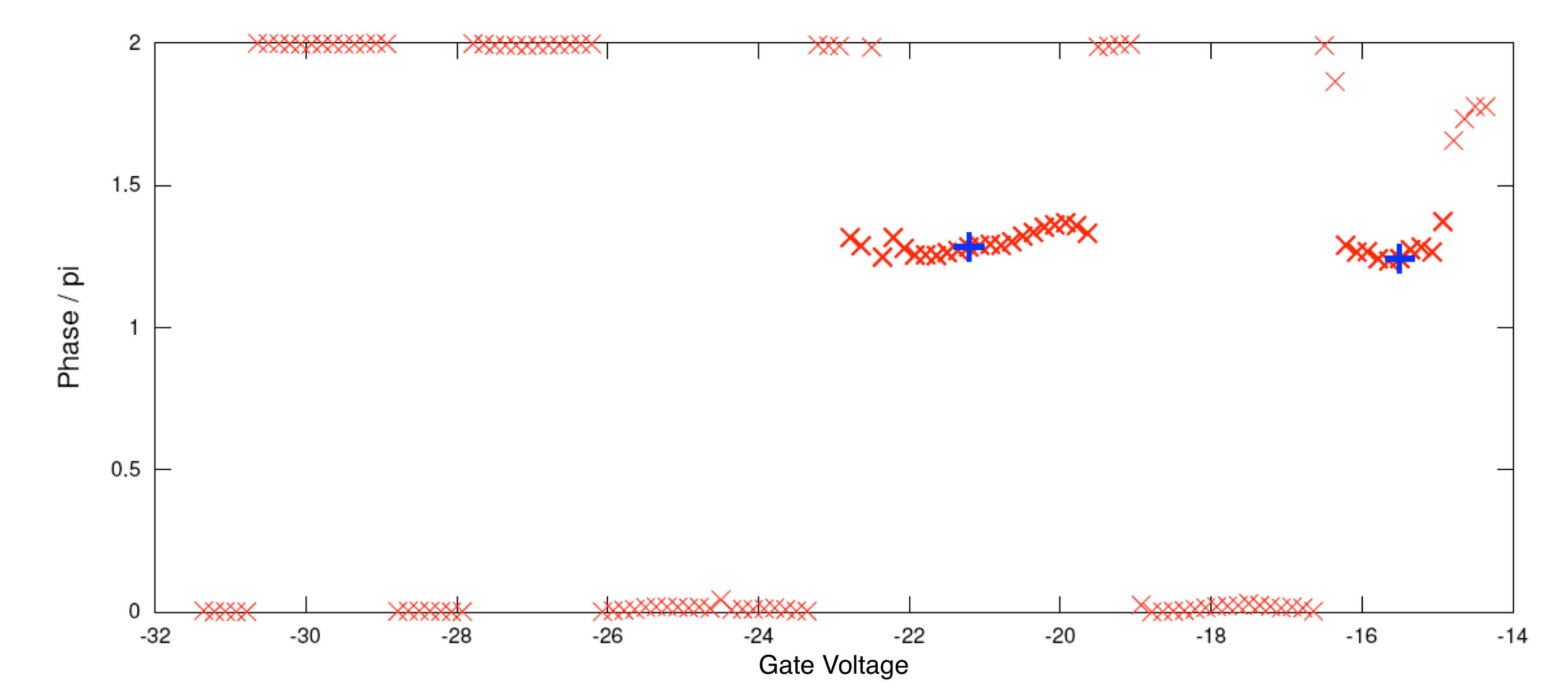}
\noindent
{\small Supplementary Figure 4: Best fit phase slip as a function of gate voltage at $\nu=7/3$.
For most values of the gate voltage the measured slip is very close to 0 or $2 \pi$ implying no phase slip is detected. However,  over two large regions phase slips are detected very close to $4 \pi/3$ (marks in bold).  The blue points are the positions in gate voltage where fits are explicitly displayed in Fig.~5 below.}

\vspace*{0.25in}

In Supplementary Fig.~4 we show the fit value of the phase slip $\phi_+ - \phi_-$ for each value of $V_P^0$ for the data taken at $\nu=7/3$.  (In the fitting program, $\delta R = .01 k\Omega$, $w = .5$mV and $\delta V = 4$ mV).   At most  values of gate voltage the measured phase slip is very close to 0 or $2 \pi$ indicating no phase slip.   However, over two regions of gate voltage the best fit clusters around the value of $4 \pi/3$ (points are bold in the figure).    The mean phase slip given by the cluster of (bold) points around $V_P = -21 mV$ is 1.30$\pi$ with a standard deviation of $0.04 \pi$.  The mean phase slip of cluster of points around -15.5mV is 1.27 $\pi$ with a standard deviation of $0.04 \pi$.   (If the slight outlier point on the right is removed from this set, then the mean goes down to 1.26 $\pi$ with standard deviation $0.02 \pi$.

To show examples of the quality of these fits, we consider the two values of $V_P^0$ marked in blue in Supplementary Fig.~4 and display the fits explicitly in Supplementary Fig.~5.
The fits for $\nu=5/2$ are made using a similar technique.  (In this case we use  $\delta R = .02 k\Omega$, $w = 1$mV and $\delta V = 7$ mV).   Plots such as that in Supplementary  Fig.~4, are made for each value of magnetic field and then these plots are re-assembled into Fig 3a of the main text.  In Fig 3a one might see three notable features.  (1) The prominent blue strip between -42 mV and -30.7mV  (2) The light magenta region between -30.7mV and -27mV, and finally (3) the yellow-green region between -27mV and -17mV --- restricted to $B$ values between approximately $B$=5.5275T and 5.5326T.    We then ``cut" these three rectangular regions out of the 2d plot 3a, and histogram all of the phase shifts in these regions, which are then shown in Fig 3b.   For the blue strip, and to a lesser extent, the yellow region, peaks are very evident.  For the magenta strip there is no clear peak in the data.

\vspace*{0.25in}

\includegraphics[width=6.15in]{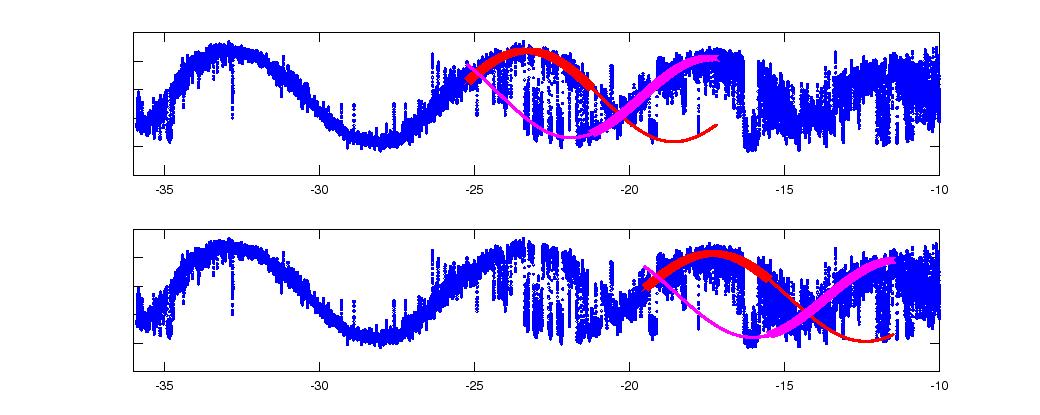}
\noindent {\small Supplementary Figure 5: Example fits.  (Top) fit for $V_P^0 = -21.2$mV (Bottom) fit for $V_P^0 = -15.8$mV.  These two sample gate voltages are marked in blue in the above figure. In both top and bottom ${\cal R}_D^-$ is marked as the thick part of the red curve, and ${\cal R}_D^+$ is the thick part of the pink curve. The point where the curves go from thick to thin is exactly $V_P^0$, where the phase slip is being measured. }

\end{document}